\documentclass[pra, aps, twocolumn, %superscriptaddress,
showpacs]{revtex4-1}

\usepackage[colorlinks,urlcolor=blue,citecolor=blue,linkcolor=blue]{hyperref}

\usepackage[english]{babel} 
\usepackage[english]{layout}
\usepackage{amsmath,amsfonts,amssymb}
\usepackage{graphicx}
\usepackage{float}
\usepackage{color}
\usepackage{braket}
\usepackage{version}
\usepackage{array,multirow,makecell}
\usepackage{afterpage}
\usepackage[toc,page]{appendix} 
\usepackage{bbold}

\renewcommand{\k}{{\bf k}}

\newcommand{\q}{{\bf q}}

\newcommand{\ob}[1]{{\color{black}#1}}
\newcommand{\rd}[1]{{\color{black}#1}}
\newcommand{\jfl}[1]{{\color{black}#1}}

\begin{document}

%\begin{comment}

\title{Polariton interactions in microcavities with atomically thin semiconductor layers}
\author{Olivier Bleu}
\affiliation{School of Physics and Astronomy, Monash University, Victoria 3800, Australia}
\affiliation{ARC Centre of Excellence in Future Low-Energy Electronics Technologies, Monash University, Victoria 3800, Australia}

\author{Guangyao Li}
\affiliation{School of Physics and Astronomy, Monash University, Victoria 3800, Australia}
\affiliation{ARC Centre of Excellence in Future Low-Energy Electronics Technologies, Monash University, Victoria 3800, Australia}

\author{Jesper Levinsen}
\affiliation{School of Physics and Astronomy, Monash University, Victoria 3800, Australia}
\affiliation{ARC Centre of Excellence in Future Low-Energy Electronics Technologies, Monash University, Victoria 3800, Australia}

\author{Meera M. Parish}
\affiliation{School of Physics and Astronomy, Monash University, Victoria 3800, Australia}
\affiliation{ARC Centre of Excellence in Future Low-Energy Electronics Technologies, Monash University, Victoria 3800, Australia}

\begin{abstract}
We investigate the interactions between exciton-polaritons in $N$ two-dimensional semiconductor layers embedded in a planar microcavity. In the limit of low-energy %and low-momentum 
scattering, where we can ignore the composite nature of the excitons, 
we obtain exact analytical expressions for the spin-triplet and spin-singlet interaction strengths, 
which go beyond the Born approximation employed in previous calculations.
Crucially, we find that the strong light-matter coupling enhances the strength of polariton-polariton interactions compared to that of the exciton-exciton interactions, %with the latter vanishing in the
%zero-momentum limit.
\ob{due to the Rabi coupling and the small photon-exciton mass ratio}.
We furthermore %show that the polariton interactions have a highly non-trivial 
\ob{obtain the dependence of the polariton interactions} on 
the number of layers 
$N$, and we highlight the important role played by the optically dark states that exist in multiple layers.
In particular, we predict that the singlet interaction strength is stronger than the triplet one for a wide range of parameters in most of the currently used transition metal dichalcogenides. This has consequences for the pursuit of polariton condensation and other interaction-driven phenomena in these materials.
\end{abstract}
\maketitle

\maketitle

\section{Introduction}

Microcavity exciton-polaritons (polaritons) are neutral quasiparticles %emerging 
that arise from the strong coupling between semiconductor excitons (bound electron-hole pairs) and cavity photon resonances.
Due to their excitonic component, polaritons interact with each other, in contrast to bare photons in vacuum. This interaction is the cornerstone of a variety of observed phenomena ranging from optical parametric scattering~\cite{Savidis2000} 
and bistability~\cite{Baas2004}, to Bose-Einstein condensation~\cite{kasprzak2006bose,balili2007bose}, superfluidity~\cite{amo2009superfluidity} and the formation of quantized vortices~\cite{lagoudakis2008quantized}. 
Hence, semiconductor microcavities are %have emerged as 
fruitful platforms to investigate two-dimensional (2D) quantum fluids of light~\cite{keeling2007collective,RMP2010polBEC,RMP2013QFL,kavokin2017microcavities}.

Atomically thin semiconductors in the form of transition metal dichalogenides (TMDs) have recently emerged as promising materials for realizing polaritonic phenomena at room temperature, due to the large exciton binding energies in TMD monolayers~\cite{He2014,ye2014probing,Chernikov2014,RMP2018TMD}.
Furthermore, TMDs can be made nearly disorder free, unlike organic materials~\cite{mikhnenko2015exciton}, and they can be externally tuned using electrostatic gating~\cite{RMP2018TMD}, which is an essential tool for any future 
%the future generation of 
optoelectronic devices~\cite{Sanvitto2016}.
Already, a strong exciton-photon (Rabi) coupling has been observed in both TMD 
monolayer~\cite{liu2015strong,flatten2016room,lundt2016room,sidler2017fermi} and multilayer structures~\cite{dufferwiel2015exciton,krol2019exciton}. 
In particular, the use of multilayer van der Waals heterostructures can generate large Rabi couplings~\cite{schneider2018two}
as well as provide routes towards engineering other material properties~\cite{geim2013van}.
However, it is an open and non-trivial question how the polariton-polariton interactions depend on experimental parameters such as the light polarization~\cite{shelykh2009polariton} and the number of layers in these systems. 
The answer to this question impacts the highly active investigation %topic 
of interaction-induced nonlinear optical properties~\cite{Scuri2018,barachati2018interacting,Tan_2020,emmanuele2019highly,gu2019enhanced} and the ongoing quest~\cite{waldherr2018observation} to realize polariton condensation in pure TMD systems.

In this manuscript, we address this question by studying the effective interactions between polaritons in a system of
$N$ identical 2D layers embedded in a planar microcavity. 
A key simplification of our work is to assume %the assumption
that the energy scale of polariton-polariton scattering is smaller than the exciton binding energy, \ob{$\varepsilon_{B}^X$,} %symbol?
\ob{thus allowing us to treat the excitons as bosons with contact interactions~\cite{Takayama2002,Schindler2008}.}
%thus allowing us to neglect the composite nature of the excitons and treat them as structureless bosons with contact interactions and mass $m_X$. 
%
This is a reasonable assumption in the case of TMD layers, where the exciton binding energy is much larger than all other relevant energy scales~\cite{RMP2018TMD,schneider2018two}.
Solving the scattering problem of two lower polaritons \ob{in the limit of} %at 
zero momentum, 
we obtain the following \textit{exact} expression  
for the polariton-polariton interaction strength:
\begin{equation}
T_{\sigma\sigma'} =\frac{4\pi \hbar^2 X_{0}^4}{m_X N \ln\left(\frac{\varepsilon_{\sigma\sigma'}}{2 |E_0^L|}\right)}\equiv \begin{cases}
    \alpha_1,& \sigma=\sigma'\\
  \alpha_2,&  \sigma\neq\sigma' .
  \end{cases}
\label{Eq.1}
\end{equation}
Here $\varepsilon_{\sigma\sigma'}>0$ are the energies associated with the spin-triplet ($\sigma=\sigma'$) and spin-singlet ($\sigma\neq\sigma'$) exciton scattering lengths, where $\sigma=\pm$ encodes the pseudo-spin (circular polarization) 
%degree of freedom 
of the exciton (photon).
$m_X$ is the exciton mass, while 
 $X^2_0$ and $E_0^L$ are, respectively, the exciton fraction and the energy (relative to the exciton energy) 
 of the zero-momentum lower polariton, 
 which depend on the number of layers $N$ via the exciton-photon 
 Rabi coupling.

\ob{Crucially, Eq.~\eqref{Eq.1} differs from the case of exciton interactions where the scattering has been shown to decrease logarithmically with collision energy~\cite{Takayama2002,Schindler2008}, as expected for 2D quantum particles with short-range interactions~\cite{adhikari1986quantum,levinsen2015strongly}.} %, as has been confirmed from microscopic exciton-exciton calculations~\cite{Takayama2002,Schindler2008}.}
Hence the strong light-matter coupling \textit{enhances} the polariton-polariton interaction strength with respect to the corresponding exciton-exciton interaction strength, \ob{which is a major qualitative difference from previous treatments based on the Born approximation~\cite{Ciuti1998,tassone1999exciton,RMP2013QFL}. 
\rd{Indeed, within the Born approximation, 
%the exciton-exciton interactions are finite and 
the coupling to light decreases the interaction strength due to the reduced exciton fraction, a feature which has become a central tenet of polariton physics~\cite{RMP2013QFL}. Our result shows that, generically, the converse is true.}
Moreover, this simple analytic expression only depends on measurable parameters and can be universally applied to a range of TMD materials \rd{and even single semiconductor quantum wells where $|E_0^L| \ll \varepsilon_{B}^X$}. 
Thus, Eq.~\eqref{Eq.1} is a key result of this work.} 
%which we derive in the following.

\rd{The paper is organized as follows. The model is introduced in Section~\ref{Sec:1} where we highlight the non-trivial subtleties of multilayer systems. In Sec.~\ref{Sec:2}, we present the derivation of the polariton-polariton scattering $T$ matrix, and explain how it differs from the standard low energy quantum scattering in 2D. Finally, we apply our results to different TMD materials and discuss the implications for experiments in Section~\ref{Sec:3}. A brief summary and our conclusions are given in Sec.~\ref{Sec:Conc}. Additional information and technical details are provided in Appendices~\ref{App:A},~\ref{App:B} and~\ref{App:C}.}

\section{Model}\label{Sec:1}
\subsection{Dark, bright and polariton states}
We start with the single-polariton %non-interacting 
Hamiltonian that describes the coupling between the cavity photon and $N$ monolayer excitonic modes:
\begin{eqnarray} \nonumber
   \hat{H}_0&=&\sum_{\mathbf{k},\sigma} E_{\mathbf{k}}^C \hat{c}_{\mathbf{k}\sigma}^{\dagger} \hat{c}_{\mathbf{k}\sigma}+ \sum_{\mathbf{k},\sigma} \sum_{n=1}^{N} E_{\mathbf{k}}^X \, \hat{x}_{\mathbf{k}\sigma, n}^{\dagger} \, \hat{x}_{\mathbf{k}\sigma, n}\\ \label{eq:ham0}
   & & + \frac{\hbar g_R}{2} \sum_{\mathbf{k},\sigma} \sum_{n=1}^{N} \left(\hat{x}_{\mathbf{k}\sigma, n}^{\dagger} \, \hat{c}_{\mathbf{k}\sigma} +\hat{c}_{\mathbf{k}\sigma}^{\dagger} \hat{x}_{\mathbf{k}\sigma, n}\right) , 
\end{eqnarray}
where $\hat{c}_{\mathbf{k}\sigma}$ ($\hat{c}_{\mathbf{k}\sigma}^\dagger$) and $\hat{x}_{\mathbf{k}\sigma,n}$ ($\hat{x}_{\mathbf{k}\sigma,n}^\dagger$) are bosonic annihilation (creation) operators of cavity photons and monolayer excitons, respectively, with in-plane momentum $\hbar\mathbf{k}$ and layer index $n$. The kinetic energies at low momenta are $E_{\mathbf{k}}^C = \hbar^2k^2/2m_C +\delta$ and  $E_{\mathbf{k}}^X = \hbar^2k^2/2m_X$, where $k \equiv |\k|$ and we measure energies with respect to the exciton energy at zero momentum.
Thus, $\delta$ is the photon-exciton detuning, while $m_C$ \ob{is the photon mass.} %and $m_X$ are the photon and exciton masses, respectively. 
Here, for simplicity, we consider identical monolayers that are located at the maxima of the electric field within the cavity, so that both $E_{\mathbf{k}}^X$ and the exciton-photon coupling $g_R$ are independent of $n$.
However, it is straightforward to generalize our results to the case of a layer-dependent light-matter coupling. 
We furthermore assume that $g_R$ is independent of polarization/spin  
and we neglect any splittings between the longitudinal and transverse modes of the excitons~\cite{Mialle1993,Glazov2014,Yu2014} and photons~\cite{Panzarini1999}.
Hence, there is no spin-orbit coupling \cite{Kavokin2005,leyder2007observation,Bleu2017,lundt2019optical} in our model, but such a single-particle effect should not strongly affect the short-distance two-body scattering processes considered here. 

 \begin{figure}%[tbp] % not "pt"
   \centering
%\begin{minipage}{\columnwidth}
\includegraphics[width=\columnwidth]{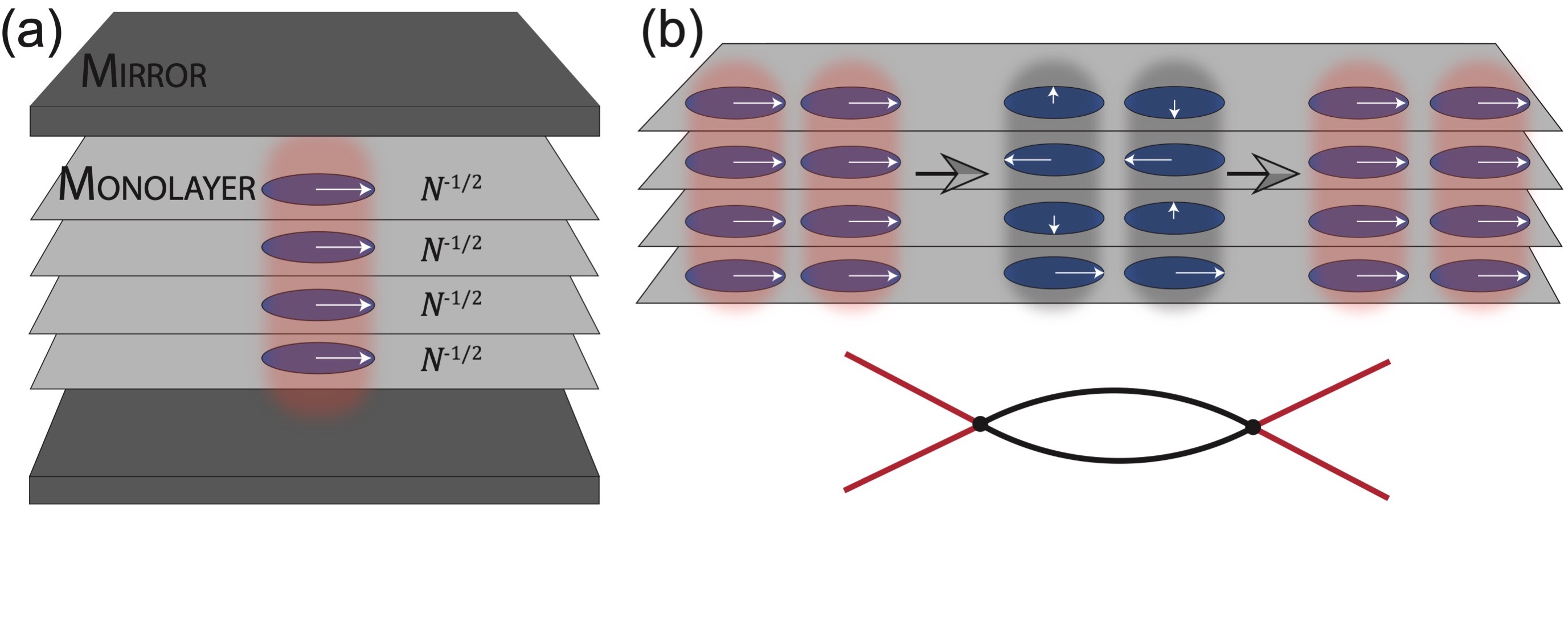}
%\end{minipage}
%\begin{minipage}{\columnwidth}
\includegraphics[width=0.9\columnwidth]{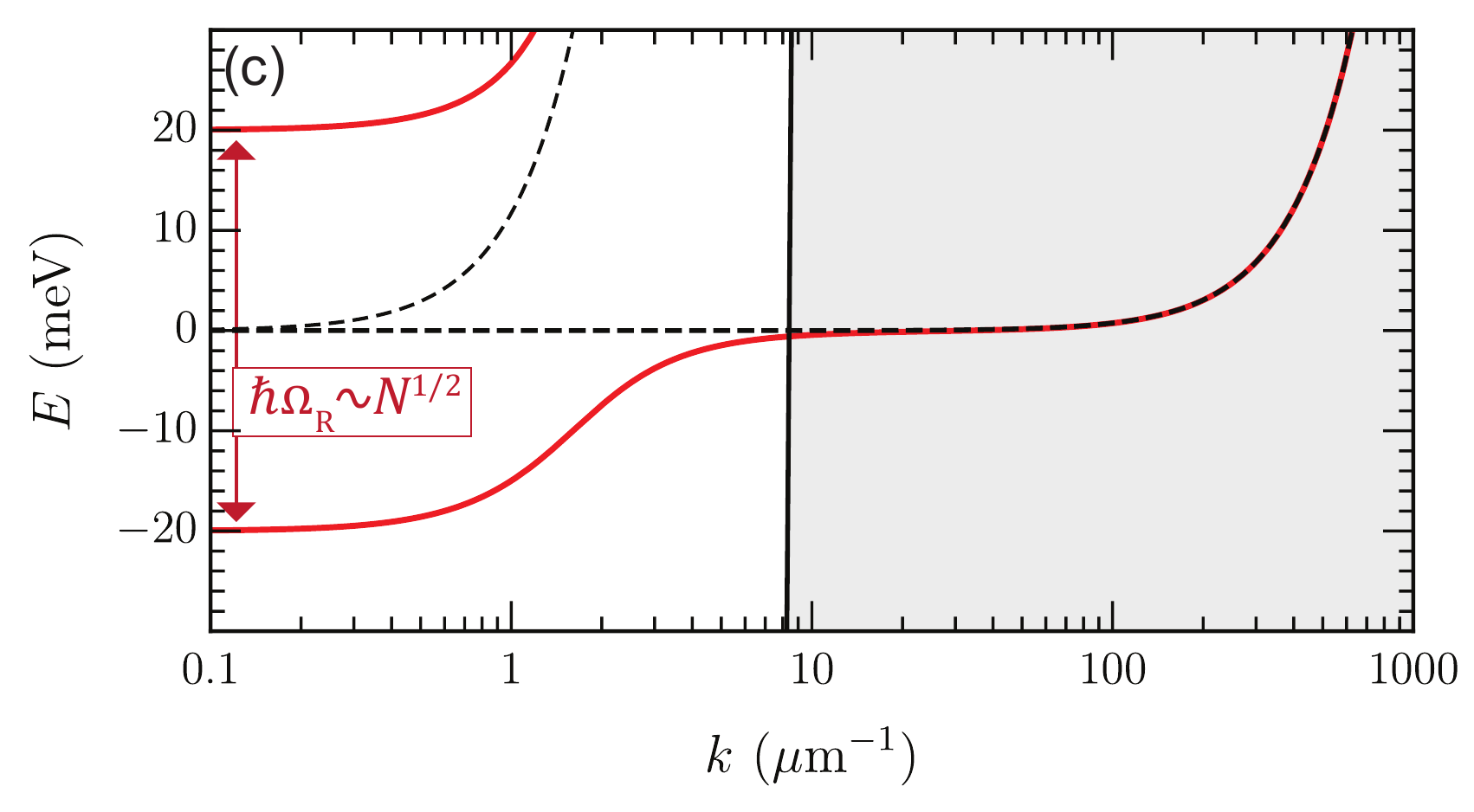}
%\includegraphics[width=0.8\columnwidth]{ExcitonVSPolariton.pdf}
%\end{minipage}
\caption{ %(Color online) 
(a) Schematic illustration of the microcavity structure with $N$ embedded monolayers, where $N=4$. A polariton consists of a cavity photon (shaded red) and a superposition of $N$ in-phase 2D excitons (blue ellipses), where the relative phase of each exciton is represented by an in-plane arrow. 
(b) Example of a scattering process involving intermediate dark states which are uncoupled to light.  
(c) Polariton dispersion (red) at zero detuning, together with the uncoupled cavity photon and exciton modes (dashed black lines). For high momenta $k > \epsilon_X/\hbar c$ (shaded region), where $\epsilon_X$ is the exciton energy and $c$ is the speed of light, the exciton is far detuned in energy from the photon and thus essentially uncoupled. 
We use the parameters for MoSe$_2$ (see Table~\ref{tab}) with %exciton energy 
$\epsilon_X = 1.66$eV~\cite{dufferwiel2015exciton} and $N=4$ layers. %(Horizontal axis in logarithmic scale) 
}
\label{fig1}
\end{figure}

The spin-degenerate eigenstates of Eq.~\eqref{eq:ham0} consist of two polariton modes (upper and lower branches) and $N-1$ \textit{dark} states which are decoupled from light~\cite{ivchenko1994resonant,kavokin2003cavity}.
The multilayer system is frequently described by a two-coupled-mode exciton-photon model with a renormalized Rabi coupling~\cite{RMP2013QFL}, but here, we keep track of the complete structure of the eigenstates \ob{which is of crucial importance when we consider polariton interactions below}.
Since only the bright states [in-phase superpositions of all monolayer excitons, as depicted in Fig.~\ref{fig1}(a)] couple to light, one can rewrite the Hamiltonian in the corresponding convenient basis:
\begin{eqnarray}
   \hat{H}_0&=&\sum_{\mathbf{k},\sigma}\left[E_{\mathbf{k}}^C \hat{c}_{\mathbf{k}\sigma}^{\dagger} \hat{c}_{\mathbf{k}\sigma}+  E_{\mathbf{k}}^X \left(\hat{b}_{\mathbf{k}\sigma}^{\dagger} \hat{b}_{\mathbf{k}\sigma}+\sum_{l=1}^{N-1} %E_{\mathbf{k}}^X
   \hat{d}_{\mathbf{k}\sigma,l}^{\dagger}\, \hat{d}_{\mathbf{k}\sigma,l}\right)\right]\nonumber \\  & & + \frac{\hbar \Omega_R}{2} \sum_{\mathbf{k},\sigma}  \left(\hat{b}_{\mathbf{k}\sigma}^{\dagger} \hat{c}_{\mathbf{k}\sigma} +\hat{c}_{\mathbf{k}\sigma}^{\dagger} \hat{b}_{\mathbf{k}\sigma}\right) ,
   \label{eq:Hbrightdark}
\end{eqnarray}
where $\hat{b}_{\mathbf{k}\sigma}$ and $\hat{d}_{\mathbf{k}\sigma,l}$ are the  annihilation operators for bright and dark states, respectively, which are related to the bare monolayer exciton operators via the unitary transformation:
\begin{equation} \label{eq:Unitary}
  \hat{d}_{\mathbf{k}\sigma,l}=\sum_{n=1}^{N} u_{ln} \, \hat{x}_{\mathbf{k}\sigma,n},
  ~~ 
  \hat{b}_{\mathbf{k}\sigma} \equiv \hat{d}_{\mathbf{k}\sigma,N} = \sum_{n=1}^{N} \frac{\hat{x}_{\mathbf{k}\sigma,n}}{\sqrt{N}} 
  %\hat{d}_{\mathbf{k}\sigma,l}^{\dagger}=\sum_{n=1}^{N}  u_{ln}^* \hat{x}_{\mathbf{k}\sigma, n}^{\dagger} ,
\end{equation}
with $u_{ln}=\frac{1}{\sqrt{N}}e^{i2\pi n l/N}$
\footnote{A similar transformation has been considered in Ref.~\cite{la1998biexcitons}, but its explicit mathematical form was not given.}.
The multilayer nature of the bright states gives rise to an enhanced Rabi coupling $\hbar \Omega_R=\hbar g_R \sqrt{N}$, thus making it easier to access the strong-coupling regime in a multilayer structure. \rd{We emphasize that the present dark states consist of superpositions of monolayer bright excitons. Thus, they should not be confused with spin-forbidden dark excitons which can exist in free monolayers and have a different spectral energy \cite{Zhang2015,Wang2017}.}

\rd{The decomposition of the Hamiltonian into the basis of dark and bright excitons in Eq.~\eqref{eq:Hbrightdark} allows us to arrive at the} diagonal form of the exciton-photon Hamiltonian:
\begin{equation} \notag
 \hat{H}_0=\! \sum_{\mathbf{k},\sigma}\! \left[E_{\mathbf{k}}^L \hat{L}_{\mathbf{k}\sigma}^{\dagger} \hat{L}_{\mathbf{k}\sigma}+ E_{\mathbf{k}}^U \hat{U}_{\mathbf{k}\sigma}^{\dagger}\hat{U}_{\mathbf{k}\sigma}+\!\! \sum_{l=1}^{N-1}  E_{\mathbf{k}}^X\hat{d}_{\mathbf{k}\sigma,l}^{\dagger}\hat{d}_{\mathbf{k}\sigma,l}\right] \! ,
\end{equation}
with $\hat{L}$ ($\hat{U}$) the lower (upper) polariton annihilation operators defined in the standard way
\begin{equation}
 \begin{pmatrix}  \hat{L}_{\mathbf{k}\sigma}\\ \hat{U}_{\mathbf{k}\sigma}  \end{pmatrix} = \begin{pmatrix}  X_{\mathbf{k}} && C_{\mathbf{k}}\\ -C_{\mathbf{k}} &&  X_{\mathbf{k}} \end{pmatrix}     \begin{pmatrix}  \hat{b}_{\mathbf{k}\sigma}\\ \hat{c}_{\mathbf{k}\sigma}  \end{pmatrix} .
\end{equation}
Here  $E_\mathbf{k}^{U,L}$ are the polariton eigen-energies [see Fig.~\ref{fig1}(c)],
\begin{align}
E_\mathbf{k}^{U,L}=\frac{1}{2} \left(E_{\mathbf{k}}^X+E_{\mathbf{k}}^C \pm \sqrt{\left(E_{\mathbf{k}}^C-E_{\mathbf{k}}^X\right)^2+ \hbar^2\Omega_R^2 }\right) ,
\end{align}
and $X_{\mathbf{k}},C_{\mathbf{k}}$ are the Hopfield coefficients, corresponding to exciton and photon fractions:
\begin{align}
X_{\mathbf{k}}^2=\frac{1}{2} \left(1+ \frac{E_{\mathbf{k}}^C-E_{\mathbf{k}}^X}{E_{\mathbf{k}}^U-E_{\mathbf{k}}^L}\right) ,~~~ C_{\mathbf{k}}^2=1- X_{\mathbf{k}}^2 .
\end{align}

\subsection{Exciton-exciton interactions} Since the layer spacing is typically larger than the exciton size, we may assume that the interactions between excitons only occur within the same layer. Furthermore, if the scattering energy is small compared to the exciton binding energy (as is the case in TMDs~\cite{RMP2018TMD}), then we can describe the exciton interactions with an $s$-wave contact potential, 
\begin{align}
%\hat{H}_{XX}
\hat{V}= \sum_{n=1}^{N}  \, \sum_{\substack{\k,\k',\q\\ \sigma,\sigma'}} \frac{ g_{\sigma\sigma'}}{2}  \hat{x}_{\mathbf{k+q}\sigma,n}^{\dagger} \hat{x}_{\mathbf{k'-q}\sigma',n}^{\dagger}  \hat{x}_{\mathbf{k'}\sigma',n} \hat{x}_{\mathbf{k}\sigma,n} ,
\end{align}
\ob{since at large separation the excitons have van der Waals interactions, which are short range~\cite{landau2013quantum}.}
%where 
The %short-range (contact) exciton-exciton 
``bare'' spin-dependent coupling strength $g_{\sigma\sigma'}$ is independent of layer index $n$ since we have assumed that the monolayers 
are identical. Also, we have set the monolayer area to 1.
%
%We emphasize that 
\ob{Our} approach is different from determining the exciton-exciton interaction strength within the Born approximation, as in previous  works~\cite{Ciuti1998,tassone1999exciton,Rochat2000,combescot2008many,Glazov2009,shahnazaryan2017exciton,Estrecho2019}. That approximation effectively estimates $g_{\sigma\sigma'}$ from the microscopic structure of the excitons, whereas here we solve the low-energy scattering problem exactly and treat $g_{\sigma\sigma'}$ as a bare parameter that must be related to experimental observables~\cite{mora2009ground}. As such, we impose a cutoff $\Lambda$ on the relative scattering momentum, which we will send to infinity at the end of the calculation. 

Transforming to the bright-dark exciton basis using
Eq.~\eqref{eq:Unitary}, the interaction term becomes:
\begin{equation}\label{eq:interNB} 
%\hat{H}^{XX}
\hat{V}= \sum_{\{l_j\}}\delta_{\mathcal{M}} \! \! \sum_{\substack{\k,\k',\q\\ \sigma,\sigma'}}\frac{ g_{\sigma\sigma'}}{2N} \hat{d}_{\mathbf{k+q}\sigma,l_1}^{\dagger}\hat{d}_{\mathbf{k'-q}\sigma',l_2}^{\dagger} \hat{d}_{\mathbf{k'}\sigma',l_3}\hat{d}_{\mathbf{k}\sigma,l_4} ,
\end{equation}
where  $\{l_j\}=\{l_1,l_2,l_3,l_4\}$. The kronecker delta ($\delta_{\mathcal{M}}=1 $ if $\mathcal{M}=0$, $\delta_{\mathcal{M}}=0 $ otherwise) encodes the phase selection rule for binary scatterings illustrated in Fig.~\ref{fig1}(b), where $\mathcal{M}=\text{Mod}{[l_1+ l_2- l_3 -l_4,N}]$.
It is worth noting that the interaction coupling constant is %renormalized 
reduced by the factor of $1/N$ in the new bright-dark-states basis. Moreover, written in this form, Eq.~\eqref{eq:interNB} can involve a huge number of terms ($N^3$ for each spin channel). This highlights the complexity of the scattering processes which can occur in any multilayer structure in the strong-coupling regime.

\begin{figure*}[htbp] % not "pt"
   \includegraphics[width=\linewidth]{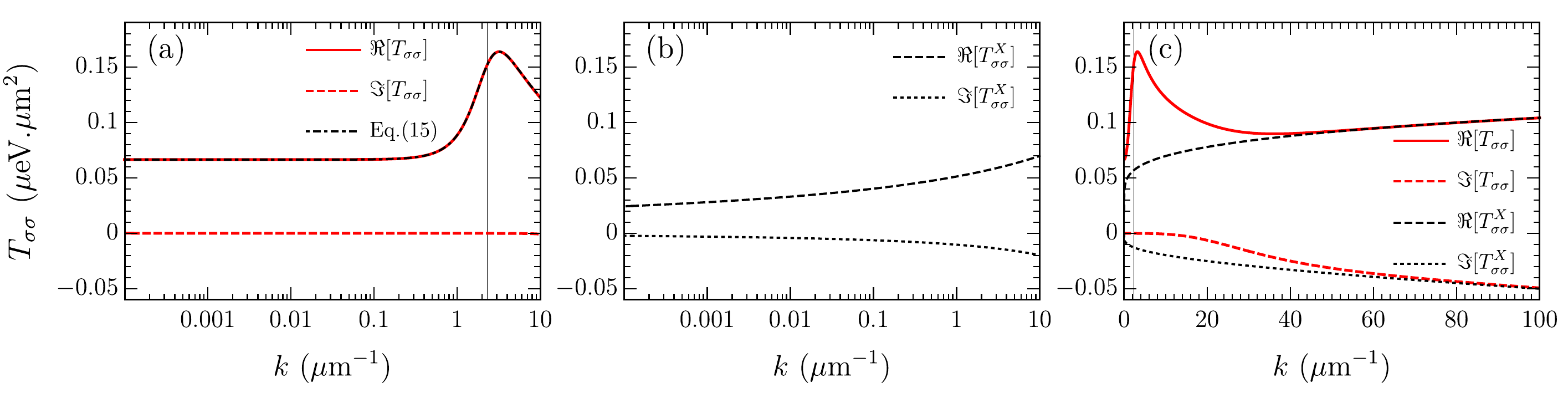}
\caption{%(Color online) 
\jfl{Polariton and exciton triplet elastic scattering as a function of relative momentum. (a) Real and imaginary parts of the full polariton $T$ matrix, Eq.~\eqref{eq:scatt1}, together with the analytical formula in the small photon mass approximation, Eq.~\eqref{eq:Tfinitek}.  (b) Low-energy exciton triplet $T$ matrix, $T_{\sigma\sigma}^X(k)$, Eq.~\eqref{eq:Tmatx}.
(c) Polariton and exciton $T$ matrices on a larger momentum scale. We use the MoSe$_2$ parameters with $m_X=1.14m_0$ \cite{HannuPekka2015}, $m_C=10^{-5}m_0$ \cite{Tan_2020}, $\delta=0$, and $N=1$, giving the inflection wave vector $q_0\simeq2.29\mu$m$^{-1}$ [thin vertical line in panels (a,c)].}}
\label{fig2}
\end{figure*}

\section{Polariton-polariton scattering}\label{Sec:2}
To investigate polariton-polariton interations, we consider the two-body scattering problem at zero center-of-mass momentum. The two-particle states are 
$\ket{A_{\sigma},B_{\sigma'},\mathbf{k}}=%\frac{1}{\sqrt{2}}
\hat{A}_{-\mathbf{k}\sigma}^{\dagger}
\hat{B}_{\mathbf{k}\sigma'}^{\dagger}\ket{0}$,
where the operators $\hat{A}$, $\hat{B}$ can correspond to lower polaritons $\hat{L}$, upper polaritons $\hat{U}$ or dark-exciton operators $\hat{d}_l$, with $l = 1, 2,  \ldots , N-1$. 
To proceed, we employ the $T$-matrix operator, which is given by the Born series
\begin{equation}
\hat{T}(E)= \hat{V} + \hat{V} \frac{1}{E-\hat{H}_0 +i0}\hat{V}+ ... 
\end{equation}
where $E$ is the scattering energy and $+i0$ represents an infinitesimal positive imaginary part. 
The interaction strength for lower polaritons is then given by the matrix element
\begin{align}
T_{\sigma\sigma'}(k) & \equiv \frac{\bra{L_{\sigma},L_{\sigma'},\mathbf{k}'}\hat{T}\!\left(2E_\k^L\right)\ket{L_{\sigma},L_{\sigma'},\mathbf{k}}}{1+\delta_{\sigma\sigma'}}, 
\end{align}
with the on-shell condition $|\mathbf{k}'|=|\mathbf{k}|=k$.
Here the normalization factor in the denominator accounts for scattering between identical particles.
The Born approximation to the interaction strength corresponds to keeping only the first term in the series, which gives %interaction strength 
$g_{\sigma \sigma'} X^4_\k/N$.
However, higher order terms will significantly modify this result %are more complicated 
since they can involve scattering into dark intermediate states, as illustrated in Fig.~\ref{fig1}(b).

Remarkably, we find that 
the low-energy polariton $T$ matrix takes the simple form (see Appendix~\ref{App:A} for the detailed calculation)
\begin{align}\label{eq:scatt1}
T_{\sigma\sigma'}(k)
&=\frac{X_{\mathbf{k}}^4}{\frac{N %\mathcal{A}
}{g_{\sigma\sigma'}} 
-\Pi\left(2E_\k^L\right)} . 
\end{align}
Here, the one-loop polarization bubble $\Pi(E)$ is extremely well approximated by that of $N$ exciton pairs:
\begin{equation}\label{eq:PiApprox}
\Pi(E)\simeq N\sum_\mathbf{q}^{\Lambda}\frac{1}{E- 2E_\mathbf{q}^{X}+i0} ,
\end{equation}
since the exciton scattering is dominated by large momenta where the photon is far off resonant [see Fig.~\ref{fig1}(c)]. 
%This is a consequence of the photon mass being orders of magnitude smaller than the exciton mass, $m_C \ll m_X$. 
\ob{This is a consequence of the small photon-exciton mass ratio, $m_C/m_X\simeq10^{-5}$ (See Appendix~\ref{App:B})}. 
%\ob{Note that, had we dropped the dark states in Eq.~\eqref{eq:ham0} as is commonly done~\cite{RMP2013QFL}, we would not have arrived at the prefactor $N$. Hence the dark states play an essential role in the interactions.}

To obtain cutoff-independent results, we relate the bare couplings to physical observables as follows~\cite{levinsen2015strongly}
\begin{equation}\label{eq:remorm}
\frac{1}{g_{\sigma\sigma'}}=-\sum_\mathbf{q}^{\Lambda}\frac{1}{\varepsilon_{\sigma\sigma'}+ 2E_\mathbf{q}^{X}} , 
\end{equation}
where we have introduced the physical energy scales
$\varepsilon_{\sigma\sigma'}=\frac{\hbar^2}{2m_r a_{\sigma\sigma'}^2}$ %,
related to the 2D exciton $s$-wave scattering lengths $a_{\sigma\sigma'}$ and the two-exciton reduced mass $m_r=m_X/2$. Note that the scattering parameters are intrinsic to the monolayer and are independent of $N$.
In the singlet case, $\varepsilon_{+-}=\varepsilon_{-+}=\varepsilon_{B}^{XX}$ corresponds to the binding energy of the biexciton (bound state of two excitons). Due to Pauli exclusion, there is no triplet biexciton state,  %bound state %associated with $E_{\sigma\sigma}$ 
but the triplet scattering length is well defined and is of the order of the 2D exciton Bohr radius $a_{\sigma\sigma}\sim a_B$ (See Appendix~\ref{App:C}); hence  we have \ob{$\varepsilon_{\sigma\sigma}\sim  \varepsilon_{B}^X$}.

Inserting Eq.~\eqref{eq:remorm} into Eq.~\eqref{eq:scatt1} and taking the limit 
 $\Lambda\rightarrow \infty$,
one obtains the cutoff-independent $T$ matrix
\begin{equation} \label{eq:Tfinitek}
%\bra{L_{\sigma},L_{\sigma'},\mathbf{k}'}\hat{T}\ket{L_{\sigma},L_{\sigma'},\mathbf{k}}
T_{\sigma\sigma'}(k)
=\frac{4\pi \hbar^2 X_{\mathbf{k}}^4}{m_X N \ln\left(\frac{\varepsilon_{\sigma\sigma'}}{2|E_{\mathbf{k}}^L|}\right)} .
\end{equation}
The limit $k \rightarrow 0$ finally yields $T_{\sigma\sigma'}$ in Eq.~\eqref{Eq.1}, which gives the lower polariton effective interaction ``constants'' for the triplet ($\alpha_1$) and singlet ($\alpha_2$) channels. %at zero momentum.  (a similar procedure can be used to evaluate the scattering between upper polaritons).
\begin{comment}
\ob{Equations~\eqref{Eq.1} and~\eqref{eq:Tfinitek} 
show that the low-momentum polariton interaction strength is enhanced compared to that of monolayer excitons because the scattering energy is dominated by the \jfl{strong light-matter} coupling --- for a detailed comparison, see the subsections~\ref{subSec:vanishing} and~\ref{subSec:Comparison} below. This is in sharp contrast to the behavior predicted by the Born approximation~\cite{Ciuti1998,tassone1999exciton,RMP2013QFL}.
Furthermore, Eq.~\eqref{Eq.1} implies that the chemical potential of a polariton condensate will scale linearly with density. This is consistent with recent measurements of the polariton blueshift in the Thomas Fermi regime~\cite{Estrecho2019}, \ob{and differs from}
%This is in contrast to 
the behavior of a 2D exciton condensate, which will feature a logarithmic dependence on density according to Bogoliubov theory~\cite{popov1972theory,mora2009ground}.}
\end{comment}
 
\jfl{
\subsection{Comparison between exciton-exciton and polariton-polariton scattering}
\label{subSec:Comparison}

Equations~\eqref{Eq.1} and~\eqref{eq:Tfinitek} are key results of this work, since they
imply that the low-momentum polariton interaction strength is enhanced compared to that of monolayer excitons,  % --- for a detailed comparison, see the subsections~\ref{subSec:vanishing} and~\ref{subSec:Comparison} below. 
which is in sharp contrast to the behavior predicted by the Born approximation~\cite{Ciuti1998,tassone1999exciton,RMP2013QFL}.
To see this, note that in the absence of light-matter coupling, %Eq.~\eqref{eq:Tfinitek} 
Eq.~\eqref{eq:Tfinitek} reduces to the usual 2D two-body $T$ matrix for quantum particles in a monolayer~\cite{adhikari1986quantum},
\begin{align} \label{eq:Tmatx}
T^X_{\sigma \sigma'}(k) = \frac{4\pi \hbar^2 }{m_X} \frac{1}{\ln\left(\frac{\varepsilon_{\sigma\sigma'}}{2E_{\mathbf{k}}^X}\right) + i\pi} ,
\end{align}
which describes the interactions of excitons. % We see that this vanishes logarithmically in the limit of zero momentum.}
In Fig.~\ref{fig2} we compare the momentum dependence of the low-energy triplet polariton and exciton $T$ matrices for parameters corresponding to a MoSe$_2$ monolayer (see Table~\ref{tab}). 
%The $x$-axis is on a log scale to highlight the difference in the low-momentum behavior. 
Crucially, we see that the strength of polariton-polariton elastic scattering, $T_{\sigma\sigma}$, is \textit{larger} than the exciton-exciton one, $T_{\sigma\sigma}^X$, despite the presence of the Hopfield factor $X_\k^4$ in Eq.~\eqref{eq:Tfinitek}. %\eqref{eq:scatt1} and~. 
By comparing Eqs.~\eqref{eq:Tfinitek} with \eqref{eq:Tmatx}, we see that this enhancement of interactions is purely driven by the difference in scattering energy, which in turn is dominated by the strong light-matter Rabi coupling.

Apart from the enhancement %of the interaction strength 
due to the strong light-matter coupling, Fig.~\ref{fig2} shows that the polariton interactions behave qualitatively differently to exciton interactions. At small momenta, the polariton interactions are strongly affected by the Hopfield-factor momentum dependence, and hence the elastic interaction strength initially increases until it reaches a maximum slightly above the inflection wave vector $q_0$ of the non-parabolic polariton dispersion. 
%Panel~(c) shows the momentum dependence of both the polariton and exciton $T$ matrices on a larger scale. % (linear $x$-axis). 
%Here, we see that 
At large wave vectors, the polariton scattering \textit{decreases}, until it becomes exciton-like and recovers the standard behavior of 2D scattering, as shown in Fig.~\ref{fig2}(c).

% because the scattering energy is dominated by the \jfl{strong light-matter} coupling.

We emphasize that the peak in the polariton scattering $T$ matrix at finite relative momentum is not due to so-called optical parametric scattering. The present maximum is for scattering at zero total momentum whereas optical parametric scattering requires a finite total momentum $\sim 2 q_0$.  

For completeness, in Fig.~\ref{fig2}(a) we display the results from the exact expression for polariton-polariton scattering, Eq.~\eqref{eq:scatt1}, as well as the analytic approximation in Eq.~\eqref{eq:Tfinitek} which relied on the fact that $m_C/m_X\ll1$ (dot-dashed black line). We see that these perfectly match for the relevant wave vectors probed in optical experiments, thus proving the validity of our approximations.

\afterpage{%    % defer execution until the next page break occurs anyway
\begin{table}[t!]
 \caption{\label{tab} Experimental values used in Figs.~\ref{fig2} and~\ref{fig3} for the monolayer Rabi splitting ($\hbar g_R$), exciton ($\varepsilon_{B}^{X}$) and biexciton ($\varepsilon_{B}^{XX}$) binding energies, all in units of meV.}
\begin{ruledtabular}
\begin{tabular}{lllll}
\textrm{Material}&
\textrm{ MoSe$_2$}&\textrm{WSe$_2$}&\textrm{MoS$_2$}&
\textrm{ WS$_2$ }\\
\colrule
 $\hbar g_R$  & 20~\cite{dufferwiel2015exciton} & 23.5~\cite{lundt2016room} & 46~\cite{liu2015strong} & 70~\cite{flatten2016room} \\
 $\varepsilon_{B}^{X}$  & 470~\cite{dufferwiel2015exciton} & 370~\cite{He2014,you2015observation} & 960~\cite{liu2015strong} & 700~\cite{flatten2016room}\\
   $\varepsilon_{B}^{XX}$  & 20~\cite{hao2017neutral} & 52~\cite{you2015observation} & 70~\cite{mai2014many} & 53~\cite{Nagler2018} \\
\end{tabular}
\end{ruledtabular}
\end{table}
}

\subsection{Polariton interactions at low momentum}\label{subSec:vanishing}
Equation~\eqref{eq:Tmatx} indicates that the exciton interactions vanish logarithmically in the limit of zero momentum. By contrast, this behavior is absent for the polariton interactions shown in Fig.~\ref{fig2}. 
%and in particular $T_{\sigma\sigma}$ remains finite as $k\to 0$.
% We see that this vanishes logarithmically in the limit of zero momentum.
However, %like the exciton $T$ matrix, % in Eq.~\eqref{eq:Tmatx}, 
in principle the 2D polariton $T$ matrix must vanish in the limit of strictly zero momentum, as is the case for any scattering of 2D quantum particles with short-range interactions. As we now explain, this strong qualitative difference between exciton and polariton interactions at very small momentum arises from the large exciton-photon mass ratio, since this implies that the typical momentum at which the polariton interactions start to approach zero %such behavior starts to be relevant
is not resolvable in realistic experiments.

We obtained the analytical expression~\eqref{eq:Tfinitek} by using the approximation~\eqref{eq:PiApprox}, which amounts to replacing the polariton one-loop polarization bubble by the exciton one, as justified in Appendix \ref{App:B}.
From Eq.~\eqref{eq24} with $E=2 E_\k^L$ one can see that the leading correction to Eq.~\eqref{eq:PiApprox} at small momentum comes from the first term in the bracket, since this diverges in the zero-momentum limit:
\begin{equation}
\mathcal{A}=\frac{m_L}{4\pi\hbar^2}X_0^4\ln{\left( \frac{2E_0^L-E}{-E}\right)},
\end{equation}
where $m_L=m_C/C_0^2$ is the lower polariton effective mass.
Keeping this term, the $T$ matrix takes the form:
\begin{eqnarray}\label{eq:lowkT}
T_{\sigma\sigma'}(k\rightarrow 0)
&\simeq& \frac{X_{\mathbf{k}}^4}{\frac{N %\mathcal{A}
}{g_{\sigma\sigma'}} 
-N\Pi_X\left(2E_\k^L\right)-\mathcal{A}}\\ \nonumber & =& \frac{4 \pi \hbar^2 X_{\mathbf{k}}^4}{m_X N \ln{\left( \frac{\varepsilon_{\sigma\sigma'}}{-2 E_\k^L}\right)-m_L\ln{\left( \frac{2E_0^L-2 E_\k^L}{-2 E_\k^L} \right) X_0^4}}},
\end{eqnarray}
and indeed vanishes as 
\begin{equation}
T_{\sigma\sigma'} \xrightarrow[k\rightarrow 0]{}\frac{4\pi \hbar^2}{m_L\ln{\left(\frac{2|E_0^L|m_L}{-\hbar^2k^2}\right)}}. 
\end{equation}
One can estimate the wave vector at which this logarithmic behavior starts to dominate by comparing the two terms in the denominator in~\eqref{eq:lowkT}.
Using the fact that $\ln{\left( \frac{\varepsilon_{\sigma\sigma'}}{2| E_0^L|}\right)}\sim1 $, the second term becomes relevant when
\begin{equation}
k^*a_{\sigma\sigma'}\sim \sqrt{\frac{m_L}{m_X}}e^{-\frac{N m_X}{2m_LX_0^4}}\sim e^{-\frac{N m_X}{m_C}},
\end{equation}
where we have dropped the prefactor and removed the Hopfield coefficients of order 1 in the last term. Because of the large exciton-photon mass ratio, $k^*$ is extremely small. For example for $m_X/m_C=10^{4}$ and $N=1$, one gets $1/k^*\sim 10^{4343}a_{\sigma\sigma'}$, corresponding to a length scale much larger than the observable universe radius ($\sim %l_{u}=
4.4 \times10^{26}$m)!
This explains why the vanishing of 2D polariton scattering at low momentum is unobservable in any experiment.}

\rd{It is worth noting that the above estimate strongly differs from the case of binary collisions between ultracold bosonic atoms in quasi-2D geometries \cite{Petrov2001}, for which the logarithmic behavior, albeit challenging to probe experimentally, is not physically impossible to reach.}

\section{Implications for experiments}\label{Sec:3}
\begin{figure*}[tbp] % not "pt"
   \includegraphics[width=\linewidth]{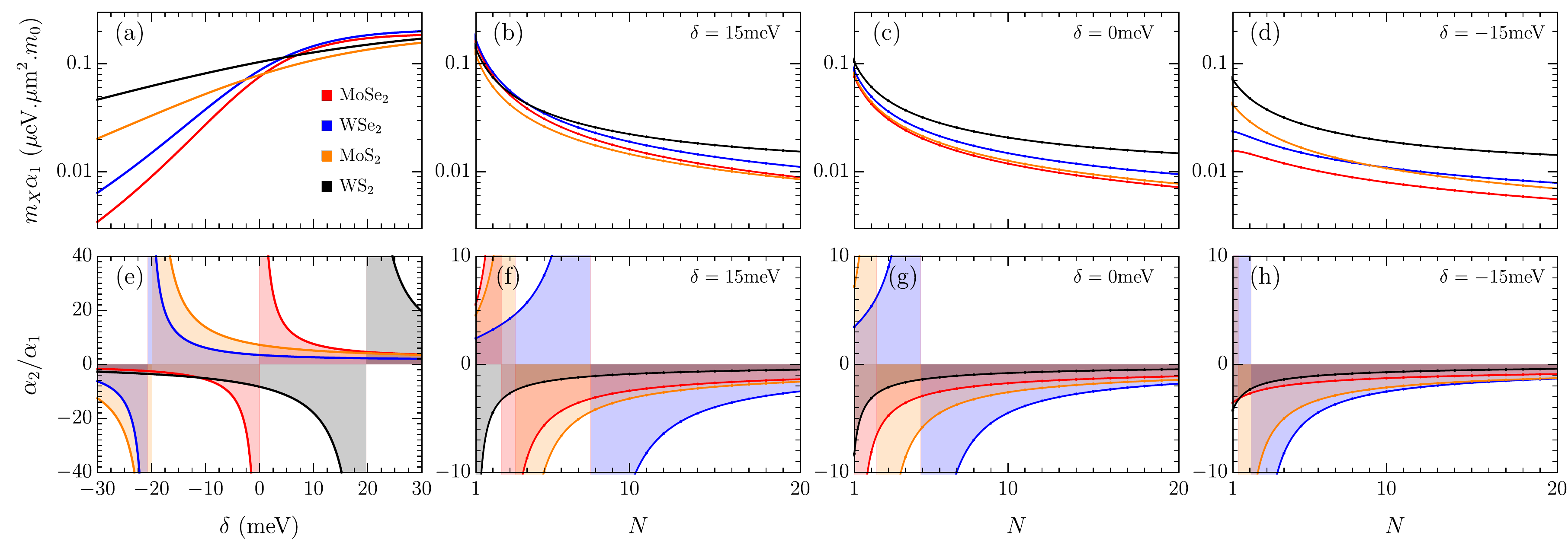}
\caption{%(Color online) 
Polariton interactions in several 2D materials. (a,e) Monolayer triplet interactions $\alpha_1$ and the corresponding singlet-triplet ratio $\alpha_2/\alpha_1$ as a function of the exciton-photon detuning $\delta$.
%(e) shows the corresponding singlet/triplet ratios.  
(b-d) Multilayer triplet interactions as a function of $N$ for different $\delta$, and (f-h) the corresponding singlet-triplet ratio. Parameters are taken from Table~\ref{tab}, with $m_0$ the free electron mass.}
\label{fig3}
\end{figure*}

\ob{We expect Eq.~\eqref{Eq.1} to be accurate for TMD layers due to the sizeable exciton binding energies %, where $\varepsilon_{B}^{X}\gtrsim 10 \hbar g_R$ 
(see Table~\ref{tab}) which imply that $|E_0^L|\ll \varepsilon_{B}^{X}$.}
Figure~\ref{fig3}(a-d) show the polariton triplet interaction strength $\alpha_1$ for a range of photon-exciton detunings $\delta$ in different 2D TMD systems.
Similar to previous predictions within the Born approximation, we find that $\alpha_1$ is repulsive and increases with increasing $\delta$ (corresponding to an increasing exciton fraction $X_0^2$).
However, we see that a larger number of layers $N$ typically suppresses the polariton \ob{triplet} interaction strength by a factor of $\sim N$ plus logarithmic corrections, \ob{suggesting}
%This suggests 
that the strongest % polariton 
interactions occur in monolayer TMDs, while the largest Rabi coupling is achieved with multiple layers. 
For the case of monolayer MoSe$_2$, the value of $\alpha_1$ at detuning $\delta\simeq-15$meV is consistent with the recent low-density measurement reported in Ref.~\cite{Tan_2020} \ob{($\alpha_1\simeq0.01 \mu$eV.$\mu$m$^2$)}, thus confirming
the validity of our model.

In contrast to the triplet case, the polariton singlet interaction strength $\alpha_2$ can display a scattering resonance when $2|E_0^L|\approx \varepsilon_{+-}$, corresponding to the point where the lower polariton branch crosses the biexciton energy.
Such resonances are present in Fig.~\ref{fig3}(e-h), where we have plotted the singlet/triplet ratio ($\alpha_2/\alpha_1$) for each TMD system.  
Here we see that the \ob{magnitude and even the sign} of $\alpha_2$ can be tuned by varying $\delta$ and/or $N$.
Furthermore, these panels demonstrate that the singlet interaction is in general stronger than the triplet one for a wide range of experimental parameters. %$N$. 
To our knowledge, this important feature has not been noticed previously. 
In particular, a large and negative $\alpha_2/\alpha_1$ can destabilize a polariton condensate, which possibly explains why condensation has been challenging to achieve thus far.
\ob{Based on current experimental data (Table~\ref{tab}), our results in Fig.~\ref{fig3}(e-h) suggest that WSe$_2$ is the most promising candidate for achieving condensation since it is easiest to access a regime with $\alpha_2>0$.}
Furthermore, the sizeable and tunable $\alpha_2$ in TMDs opens up the possibility of realizing strongly correlated phenomena such as polariton blockade~\cite{carusotto2010feshbach}, bipolariton superfluidity~\cite{Marchetti2014} and polaron physics~\ob{\cite{LevinsenPolaron2019}}.

\rd{Thus far, we have focussed our discussion on the case of TMDs, where the large exciton binding energies mean that Eq.~\eqref{Eq.1} is immediately applicable. In conventional quantum well semiconductor microcavities, the ratio of Rabi coupling to exciton binding energy is somewhat larger. For instance, in the case of a single GaAs quantum well, the ratio $\hbar\Omega_R/\varepsilon_{B}^{X}\simeq 0.35$, and therefore we still expect that our results are reasonably accurate at a quantitative level. Equation~\eqref{Eq.1} then implies that the singlet interaction should be dominant, given the proximity to the biexciton resonance. Indeed this is consistent with experimental results, since the biexciton resonance has only been probed in single GaAs quantum well structures \cite{takemura2014polaritonic,Vladimora2010,Takemura2017}. We note that the enhancement of $\alpha_2$ in this context, known as the polariton ``Feshbach resonance'', was proposed in Ref.~\cite{Woutersresonant2007}. On the other hand, for multiple quantum wells, such as structures with 12 layers, the Rabi coupling is comparable to the exciton binding energy \cite{Estrecho2019}. While this means that we are formally outside the regime of validity of Eq.~\eqref{Eq.1}, we may still draw some qualitative conclusions based on this equation. In particular, from Eq.~\eqref{Eq.1} we see that the enhanced Rabi coupling due to the multiple quantum wells means that the biexciton resonance is off resonant, and therefore the triplet interaction should dominate. This is also consistent with experiment --- indeed in experiments with multiple quantum wells the singlet interaction is typically ignored \cite{Estrecho2019}.}

\begin{comment}
The singlet %($\alpha_2$) 
resonance effectively corresponds to
the polariton ``Feshbach resonance''~\cite{Woutersresonant2007} experimentally investigated in single GaAs quantum wells~\cite{takemura2014polaritonic,Takemura2017}. 
%
\ob{For TMDs, the resonance width is larger because of the larger biexciton binding energy; thus, it should be easier to resolve experimentally.}
%
We note that GaAs-based microcavities \rd{appear to be} qualitatively different from the TMD case since values $|\alpha_2|/\alpha_1<1$ have been reported at negative detunings~\cite{Vladimora2010}, below the biexciton resonance. \ob{Moreover, the exciton binding energy in these systems} is not necessarily the largest energy scale --- for example, for the case of 12 quantum wells, we have $\hbar \Omega_R \gtrsim \varepsilon_{B}^{X}$~\cite{Estrecho2019}. 
\ob{However, Eq.~\eqref{Eq.1} and our prediction of enhanced polariton interactions should be qualitatively applicable to quantum wells. }
%On the other hand, }
\end{comment}

Note, though, that neither TMDs nor GaAs quantum wells are expected to feature a triplet polariton Feshbach resonance since there is no biexciton in this case:
% in Eq.~\eqref{Eq.1}. % since our theory is only valid when $|E_0^L|\ll E_B^X$. Indeed, %and thus it fails before we reach the resonance condition.
%Indeed,
While Eq.~\eqref{Eq.1} naively predicts a resonance at \ob{$2|E_0^L| \simeq \varepsilon_{B}^{X}$}, such a large energy scale goes beyond the validity of our model and requires 
%for large energies $|E_0^L| \approx E_B^X$ where Eq.~\eqref{Eq.1} naively predicts a resonance, the determination of the interactions requires 
a more precise description of the short-distance physics 
%that encodes 
such as the composite nature of the excitons and the layer thickness. A full microscopic description of \ob{such details} is beyond the scope of the present work.

We emphasize that most of the intermediate states appearing in the $T$ matrix are dark states [Fig.~\ref{fig1}(b)] which lie at the exciton energy. In our calculation, these correspond to virtual states that only contribute to the final strength of polariton interactions. However, in principle they can be turned into real long-lived excitons via additional scattering processes, for example mediated by phonons~\cite{Piermarocchi1996,Savenko2013,Stepanov_2019}, and they can therefore  populate an excitonic reservoir. \rd{Furthermore, to our knowledge, spontaneous polariton condensation has never been observed in single-quantum-well microcavities, which suggests that these dark states might also play an important role in the condensation process routinely reported in multi-quantum-well structures \cite{kasprzak2006bose,balili2007bose,lagoudakis2008quantized}.}

\jfl{
Finally, the finite value of zero-momentum polariton interactions given by Eq.~\eqref{Eq.1} implies that the chemical potential of a polariton condensate will scale linearly with density. This is consistent with recent measurements of the polariton blueshift in the Thomas Fermi regime~\cite{Estrecho2019}, and differs from
%This is in contrast to 
the behavior of a 2D exciton condensate, which will feature a logarithmic dependence on density according to Bogoliubov theory~\cite{popov1972theory,mora2009ground}.}
\rd{Such a logarithmic dependence has already been observed in condensates of paired fermionic atoms in quasi-2D geometries~\cite{Boettcher2016}.}

\section{Summary and conclusions}\label{Sec:Conc}
We have derived exact analytical expressions for the polariton-polariton triplet and singlet interaction strengths in TMD layers embedded in a planar microcavity. 
Crucially, we have demonstrated that the strong exciton-photon coupling enhances the polariton interactions relative to those of bare excitons.
\rd{The present work, together with the recent demonstration of the enhancement of polariton-electron scattering within a microscopic theory %going beyond the Born approximation 
\cite{li2020enhanced,li2020theory}, suggest that such  strengthening of interactions induced by the strong light-matter coupling is a universal phenomenon in these systems, which has been unnoticed so far.}

Furthermore, we have analyzed the dependence on the number of layers and we have exposed the important role of optically dark states in multilayer polariton-polariton scattering. 
Our results suggest that the singlet interaction is stronger than the triplet one for a range of TMD heterostructures, which has important consequences for realizing polariton condensation and other interaction-driven phenomena in these systems.
In particular, a large repulsive singlet interaction can lead to the formation of spin-polarized domains, and thus to spin-resolved ultra-low threshold lasing.\\

%\paragraph*{Note added.} After submission of this manuscript, we became aware of a related work in which a similar expression was obtained for $\alpha_1$ in a monolayer system within a Bogoliubov approach aiming to describe a zero-temperature scalar polariton condensate \cite{hu2020twodimensional}.

\paragraph*{Note added in proof.} After submission of this manuscript, a related work appeared in which a zero-temperature scalar polariton condensate in a monolayer system was theoretically investigated using a Bogoliubov approach, leading to a similar expression for $\alpha_1$ \cite{hu2020twodimensional}.

\acknowledgements 
We are grateful to F. M. Marchetti, E. Ostrovskaya, and M.~Pieczarka for useful discussions.  We acknowledge support from the Australian Research Council Centre of Excellence in Future Low-Energy Electronics Technologies (CE170100039).  JL is also supported through the Australian Research Council Future Fellowship FT160100244.

%\end{comment}

%\end{document}

%\renewcommand{\theequation}{S\arabic{equation}}
%\renewcommand{\thefigure}{S\arabic{figure}}

\onecolumngrid

%\newpage

%\setcounter{equation}{0}
%\setcounter{figure}{0}
%\setcounter{page}{1}

%%%%%%%%%%%%%%%%%%%%%%%%%%%%%%%%%%%%%%%%%%%%%%%%%%%%
%\clearpage

\appendix

\begin{comment}
\section*{SUPPLEMENTAL MATERIAL: ``Polariton interactions in microcavities with atomically thin semiconductor layers''}

\begin{center}
Olivier Bleu,
Guangyao Li,
Jesper Levinsen,
Meera M. Parish\\
\emph{\small School of Physics and Astronomy, Monash University, Victoria 3800, Australia and}\\
\emph{\small ARC Centre of Excellence in Future Low-Energy Electronics Technologies, Monash University, Victoria 3800, Australia}\\

\setcounter{secnumdepth}{1}
\end{center}
\end{comment}

\section{Polariton $T$-matrix}\label{App:A}
Here, we provide some details of the $T$-matrix calculation for the scattering between two lower polaritons.
First, we recall the bosonic commutation rules:
\begin{equation}
\left[\hat{A}_{\mathbf{k}\sigma }, \hat{B}_{\mathbf{k}'\sigma' }^{\dagger}\right]=\delta_{AB}\delta_{\sigma\sigma' }\delta_{\mathbf{k}\mathbf{k}' },
\end{equation}
where $\hat{A}$, $\hat{B}$ can correspond to lower polaritons $\hat{L}$, upper polaritons $\hat{U}$ or dark-exciton operators $\hat{d}_l$, with $l = 1, 2,  \ldots , N-1$. 
The two-particle states with zero total momentum are defined as: 
\begin{equation}
\ket{A_{\sigma},B_{\sigma'},\mathbf{k}}=\hat{A}_{-\mathbf{k}\sigma }^{\dagger}\hat{B}_{\mathbf{k}\sigma' }^{\dagger}\ket{0},
\end{equation}
with the scalar product:
\begin{eqnarray}
\braket{A_{\sigma_1},B_{\sigma_1'},\mathbf{k}_1|C_{\sigma_2},D_{\sigma_2'},\mathbf{k}_2}&=&\bra{0}\hat{B}_{\mathbf{k}_1\sigma_1'} \hat{A}_{-\mathbf{k}_1\sigma_1  }\hat{C}_{-\mathbf{k}_2\sigma_2  }^{\dagger}\hat{D}_{\mathbf{k}_2\sigma_2' }^{\dagger}\ket{0}\\
&=&\delta_{BC}\delta_{AD}\delta_{\sigma_1'\sigma_2 }\delta_{\sigma_1\sigma_2' }\delta_{\mathbf{k}_1,-\mathbf{k}_2 }+\delta_{BD}\delta_{AC}\delta_{\sigma_1\sigma_2}\delta_{\sigma_1'\sigma_2' }\delta_{\mathbf{k}_1\mathbf{k}_2 } .
\end{eqnarray}
%Hence, $\braket{A_{\sigma},A_{\sigma},0|A_{\sigma},A_{\sigma},0}=1$ and $\braket{A_{\sigma},A_{\sigma},\mathbf{k}|A_{\sigma},A_{\sigma},\pm\mathbf{k}}=1/2$  (if $\mathbf{k}\neq 0$). 
The non-interacting eigenvalues are given by 
\begin{eqnarray}
 \hat{H}_0\ket{A_{\sigma},B_{\sigma'},\mathbf{k}}&=&\sum_{J}\sum_{\mathbf{q},s}E_{\mathbf{q}}^J \hat{J}_{\mathbf{q}s}^{\dagger} \hat{J}_{\mathbf{q}s} \hat{A}_{-\mathbf{k}\sigma }^{\dagger}\hat{B}_{\mathbf{k}\sigma' }^{\dagger}\ket{0}\\
&=&\left( E_{\mathbf{-k}}^A+E_{\mathbf{k}}^B \right)\hat{A}_{-\mathbf{k}\sigma }^{\dagger}\hat{B}_{\mathbf{k}\sigma' }^{\dagger}\ket{0} \\
&=&\left( E_{\mathbf{-k}}^A+E_{\mathbf{k}}^B \right)\ket{A_{\sigma},B_{\sigma'},\mathbf{k}},
\end{eqnarray}
where the sum on $J$ in the first line accounts for all the available single-particle energy states, and $E_\q^J \equiv E^X_\q$ for all the dark states with $\hat{J} = \hat{d}_l$.

Now we consider the scattering between two lower polaritons, the first term of the Born series gives:
\begin{equation}
\bra{L_{\sigma},L_{\sigma'},\mathbf{k}'}\hat{V}\ket{L_{\sigma},L_{\sigma'},\mathbf{k}}=X_{\mathbf{k}'}^2X_{\mathbf{k}}^2 \frac{ g_{\sigma \sigma'}}{N} (1+\delta_{\sigma\sigma'}).
\end{equation}
(Notice that $g_{+-}=g_{-+}$, and $g_{++}=g_{--}$).
Higher order terms in the $T$-matrix involve dark or upper polariton intermediate states. It is useful to introduce the matrix element between two lower polaritons and an arbitrary two-particle state
\begin{equation}
\bra{A_{\sigma},B_{\sigma'},\mathbf{k}'}\hat{V}\ket{L_{\sigma},L_{\sigma'},\mathbf{k}} =\frac{g_{\sigma\sigma'}X_{\mathbf{k}}^2}{N}(1+\delta_{\sigma\sigma'})\begin{cases}
  \delta_{\text{Mod}[l_1+l_2,N]} ,& A=d_{l_1},~ B=d_{l_2}, ~~ l_1,l_2 \neq N\\
  X_{\mathbf{k}'}^2,& A=L,~ B=L\\
    C_{\mathbf{k}'}^2,& A=U,~ B=U
    \\
   - X_{\mathbf{k}'}C_{\mathbf{k}'} ,& A=L,~ B=U   \\
   - X_{\mathbf{k}'}C_{\mathbf{k}'} ,& A=U,~ B=L, 
  \end{cases} 
\end{equation}
and the completeness relation
\begin{equation}
\mathbb{1}=  \frac{1}{2}\sum_{\mathbf{q}}\sum_{A,B}\sum_{s,s'}\ket{A_{s},B_{s'},\mathbf{q}}\bra{A_{s},B_{s'},\mathbf{q}},
\end{equation}
which satisfies the usual property of the identity
\begin{eqnarray}
\mathbb{1}\ket{C_{\sigma},D_{\sigma'},\mathbf{k}}&= & \frac{1}{2}\sum_{\mathbf{q}}\sum_{A,B}\sum_{s,s'}\ket{A_{s},B_{s'},\mathbf{q}}\braket{A_{s},B_{s'},\mathbf{q}|C_{\sigma},D_{\sigma'},\mathbf{k}}\\ &=&\frac{1}{2}  \sum_{\mathbf{q}}\left(\ket{C_{\sigma},D_{\sigma'},\mathbf{q}}\delta_{\mathbf{q}\mathbf{k}}+\ket{D_{\sigma'},C_{\sigma},\mathbf{q}}\delta_{\mathbf{q},-\mathbf{k}}\right) \\ &=&\frac{1}{2}\left(\ket{C_{\sigma},D_{\sigma'},\mathbf{k}}+\ket{D_{\sigma'},C_{\sigma},-\mathbf{k}}\right) \\ &=&\ket{C_{\sigma},D_{\sigma'},\mathbf{k}}.
\end{eqnarray}

Then the second-order term reads 
\begin{eqnarray}
\bra{L^\sigma,L^{\sigma'},\mathbf{k}'} \hat{V} \frac{1}{E-\hat{H}_0}\hat{V}\ket{L^\sigma,L^{\sigma'},\mathbf{k}} &=&\bra{L^\sigma,L^{\sigma'},\mathbf{k}'}\hat{V} \mathbb{1}\frac{1}{E-\hat{H}_0}\hat{V}\ket{L^\sigma,L^{\sigma'},\mathbf{k}} \\
& =&(1+\delta_{\sigma\sigma'})\left(\frac{g_{\sigma\sigma'}}{N}\right)^2X_{\mathbf{k}}^2X_{\mathbf{k}'}^2  \\&& \times\sum_{\mathbf{q}}\left[ \frac{X_{\mathbf{q}}^4}{E- 2E_\mathbf{q}^{L}}+\frac{C_{\mathbf{q}}^4}{E- 2E_\mathbf{q}^{U}}+\frac{2C_{\mathbf{q}}^2X_{\mathbf{q}}^2}{E -E_\mathbf{q}^{L}-E_\mathbf{q}^{U}}+\sum_{l_1,l_2}^{N-1}\frac{\delta_{\text{Mod}[l_1+l_2,N]}}{E- 2E_\mathbf{q}^{X}}\right]\nonumber\\&=&(1+\delta_{\sigma\sigma'}) \frac{g_{\sigma\sigma'}X_{\mathbf{k}}^4}{N}\frac{g_{\sigma\sigma'}}{N}\Pi(E),
\end{eqnarray}
where we have used $|\mathbf{k}'|=|\mathbf{k}|$ in the last line. The one-loop polarization bubble, $\Pi(E)$, is given by
\begin{eqnarray}\label{eq:Pi}
\Pi(E)&=&\underbrace{\sum_\mathbf{q}^{\Lambda}\frac{X_{\mathbf{q}}^4}{E-2 E_\mathbf{q}^{L}}+\sum_\mathbf{q}^{\Lambda}\frac{C_{\mathbf{q}}^4}{E-2 E_\mathbf{q}^{U}}+2\sum_\mathbf{q}^{\Lambda}\frac{X_{\mathbf{q}}^2C_{\mathbf{q}}^2}{E- E_\mathbf{q}^{L}-E_\mathbf{q}^{U}}}_{\Pi_P(E)}+(N-1)\underbrace{\sum_\mathbf{q}^{\Lambda}\frac{1}{E- 2E_\mathbf{q}^{X}}}_{\Pi_X(E)}, %\\&=&\Pi_P(E)+(N-1)\Pi_X(E),
\end{eqnarray}
where we have introduced the high-energy cutoff wave vector $\Lambda$. \ob{Note that, had we dropped the dark states in Eq.~\eqref{eq:ham0} as is commonly done~\cite{RMP2013QFL}, the term $\left(N-1\right)\Pi_X$ would be absent. Hence the dark states play an essential role in the interactions.}

The generalisation to higher order terms leads to
\begin{eqnarray}
\bra{L_\sigma,L_{\sigma'},\mathbf{k}'} \hat{T}(E) \ket{L_\sigma,L_{\sigma'},\mathbf{k}}&=& (1+\delta_{\sigma\sigma'})\frac{g_{\sigma\sigma'}X_{\mathbf{k}}^4}{N}\left[1+\frac{g_{\sigma\sigma'}}{N}\Pi(E)+\left(\frac{g_{\sigma\sigma'}}{N}\Pi(E)\right)^2+...\right]\\&=&(1+\delta_{\sigma\sigma'})\frac{X_{\mathbf{k}}^4}{\left(\frac{N}{g_{\sigma\sigma'}}-\Pi(E)\right)},
\end{eqnarray}
which after rearranging the prefactor, gives the formula~\eqref{eq:scatt1} presented in the main text. Note that $\Pi(E)$ also contains a factor $N$.
The integrals in $\Pi(E)$ are dominated by the large wavectors $\mathbf{q}$, where $E_\mathbf{q}^{L}\rightarrow E_\mathbf{q}^{X}$, $X_{\mathbf{q}}^2\rightarrow 1$, $C_{\mathbf{q}}^2\rightarrow 0 $ and one can neglect the second and third terms. Thus, $\Pi(E)$ simply reduces to $N \times \Pi_X(E)$. This approximation is valid because $m_C\ll m_X$ as explained below.

\section{$m_C\ll m_X$ and $\Pi(E)$ approximation}\label{App:B}

The approximation of the one-loop polarization bubble [Eq.~\eqref{eq:Pi}] relies on the very large difference between the cavity photon and the exciton masses ($m_C/m_X\sim 10^{-4}-10^{-5}$).
This approximation is equivalent to neglecting the low-$q$ behavior of the integrand in:
\begin{equation}
\Pi_P(E)=\frac{1}{2\pi}\int_0^{\Lambda}q dq \left(\frac{X_{\mathbf{q}}^4}{E-2 E_\mathbf{q}^{L}}+\frac{C_{\mathbf{q}}^4}{E-2 E_\mathbf{q}^{U}}+2\frac{X_{\mathbf{q}}^2C_{\mathbf{q}}^2}{E- E_\mathbf{q}^{L}-E_\mathbf{q}^{U}}\right).
\end{equation}
To analytically evaluate if this low-q behavior plays a role, one can approximate the integrand in two domains with the following replacements:
\begin{itemize}
    \item $q<q_0$: $E_{\mathbf{q}}^L\rightarrow  \tilde{E}_{\mathbf{q}}^L=\hbar^2q^2/2m_L+E_{0}^L$, $E_{\mathbf{q}}^U\rightarrow  \tilde{E}_{\mathbf{q}}^U=\hbar^2q^2/2m_U+E_{0}^U$, $C_{\mathbf{q}}\rightarrow  C_{0}$, $X_{\mathbf{q}}\rightarrow  X_{0}$
    \item $q>q_0$: $E_{\mathbf{q}}^L\rightarrow  E_{\mathbf{q}}^X$, $C_{\mathbf{q}}\rightarrow  0$, $X_{\mathbf{q}}\rightarrow 1$.
\end{itemize}
The inflection wave vector $q_0$ and the lower (upper) polariton effective masses $m_L$, ($m_U$) are defined as:
 \begin{equation}
q_0=\frac{\sqrt{2m_L |E_{0}^L|}}{\hbar}, ~~ m_L=\frac{m_C}{C_0^2}, ~~ m_U=\frac{m_C}{X_0^2}.
\end{equation}
This gives
\begin{eqnarray}
\Pi_P(E)&\simeq&\frac{1}{2\pi}\int_0^{q_0}q dq \left(\frac{X_0^4}{E-2 \tilde{E}_\mathbf{q}^{L}}+\frac{C_0^4}{E-2 \tilde{E}_\mathbf{q}^{U}}+2\frac{X_0^2C_0^2}{E- \tilde{E}_\mathbf{q}^{L}-\tilde{E}_\mathbf{q}^{U}}\right)+\frac{1}{2\pi}\int_{q_0}^{\Lambda}q dq \left(\frac{1}{E-2 E_\mathbf{q}^{X}}\right)\\
&=&\frac{m_C}{4\pi\hbar^2 C_0^2 X_0^2}\left[X_0^6\ln{\left( \frac{2E_0^L-E}{-E}\right)}+C_0^6\ln{\left( \frac{2E_0^U-E}{2E_0^U-E+2|E_0^L|X_0^2/C_0^2}\right)}+4X_0^4C_0^4\ln{\left( \frac{(\delta-E) C_0^2X_0^2}{C_0^2(\delta-E) +|E_0^L|}\right)} \right]\nonumber \\&&+\frac{m_X}{4\pi\hbar^2}\ln{\left( \frac{E-2|E_0^L|C_0^2m_C/m_X}{E-2E_{\Lambda}^X}\right)} \label{eq24} \\
&\simeq&\frac{m_X}{4\pi\hbar^2}\ln{\left( \frac{E}{E-2E_{\Lambda}^X}\right)}=\sum_\mathbf{q}^{\Lambda}\frac{1}{E- 2E_\mathbf{q}^{X}}. %\frac{1}{2\pi}\int_0^{\Lambda}q dq\frac{1}{E- 2E_\mathbf{q}^{X}}.
\end{eqnarray}
Hence, one can see that in the limit $m_C/m_X\rightarrow 0$, $\Pi_P(E)\rightarrow \Pi_X(E)$, and Eq.~\eqref{eq:Pi} reduces to  $\Pi(E)\simeq N\Pi_X(E)$.

\section{Triplet exciton-exciton scattering} \label{App:C}
In the main text, we assume that the exciton triplet scattering length is of the order of the exciton Bohr radius $a_B$. Here, we motivate this assumption by recalling that this is the case for 2D hard disk particles of radius $r_0$. For low-energy particles, the two-body scattering is dominated by the $s$ wave contribution, and the center of mass wavefunction obeys the 2D radial Schrödinger equation~\cite{landau2013quantum}
 \begin{equation}
-\frac{\hbar^2}{2 m_r}\frac{1}{r}\frac{\partial}{\partial r}\left(r\frac{\partial \psi}{\partial r}\right) + U(r) \psi=E\psi,
\end{equation}
with $m_r$ the two-body reduced mass. 
The hard disk potential corresponds to an infinitely high potential barrier of radius $r_0$
 \begin{equation}
U =\begin{cases}
  \infty ,~~ r\leqslant r_0  \\
  0 ,~~ r>r_0.
  \end{cases} 
\end{equation}
Outside the potential, the general solution is a superposition of first and second kind Bessel functions
 \begin{equation}
\psi(r)=A J_0(kr)+B Y_0(kr),
\end{equation}
with $k=\sqrt{2m_rE}/\hbar$.
Then, using its asymptotic form, one introduces the scattering phase shift $\delta_s$
 \begin{equation}
\psi(r)\xrightarrow[r\rightarrow \infty]{}\sqrt{\frac{2}{\pi k r}}\left(A \cos(kr-\pi/4)+B \sin(kr-\pi/4)\right)=\sqrt{\frac{2}{\pi k r}}C\cos(kr-\pi/4+\delta_s),
\end{equation}
with
 \begin{equation}
\tan(\delta_s)=-\frac{B}{A}.
\end{equation}
The continuity of the wavefunction at $r=r_0$ gives the relation:
 \begin{equation}
\cot\left(\delta_s\right)=-\frac{A}{B} =\frac{Y_0(kr_0)}{J_0(kr_0)}.
\end{equation}
Finally, taking the low-energy (low-$k$) limit one obtains
 \begin{equation}
\cot\left(\delta_s\right) = \frac{2}{\pi} \ln\left(k a_s\right),
\end{equation}
where we have introduced the 2D scattering length $a_s$
 \begin{equation}
a_s=\frac{e^{\gamma}r_0}{2}\simeq 0.89 r_0,
\end{equation}
with $\gamma=0.577...$ the Euler-Mascheroni constant.

The corresponding hard disc triplet exciton $T$-matrix reads:
\begin{equation}
   T_{\sigma\sigma}^{X}(k)= \frac{2\pi \hbar^2 }{m_r} \left[ \ln\left(\frac{\varepsilon_{\sigma\sigma}}{E}\right) + i\pi\right]^{-1},
\end{equation}
with 
\begin{equation} 
\varepsilon_{\sigma\sigma}=\frac{\hbar^2}{2m_ra_s^2}.
\end{equation}

\ob{Note that a similar form for the low-energy $T$ matrix has been obtained in 
Refs.~\cite{Takayama2002,Schindler2008}, both accounting for the composite (electron-hole) nature of the 2D excitons with different theoretical approaches.
We emphasize that the exciton-exciton scattering vanishes only at zero momentum (see Fig.~\ref{fig2} (b,c)), and an arbitrary exciton momentum distribution (such as Maxwell-Boltzmann) necessarily leads to non-zero scattering. Hence, the above simple formula does not contradict the signatures of exciton-exciton interaction reported in experiments (such as in Ref.~\cite{Moody_2015}).}

\twocolumngrid

\bibliography{main} 

%\bibliography{main} 
\end{document}